\documentclass[%
preprint,
superscriptaddress,
 amsmath,amssymb,
prl,
]{revtex4-1}

\usepackage{graphicx}

\begin{document}
\title{Connection between coherent phonons and electron-phonon coupling in Sb (111)}

\author{S. Sakamoto}
\affiliation{Stanford Institute for Materials and Energy Sciences, SLAC National Accelerator Laboratory, 2575 Sand Hill Road, Menlo Park, California 94025, USA}

\author{N. Gauthier}
\affiliation{Stanford Institute for Materials and Energy Sciences, SLAC National Accelerator Laboratory, 2575 Sand Hill Road, Menlo Park, California 94025, USA}

\author{P. S. Kirchmann}
\affiliation{Stanford Institute for Materials and Energy Sciences, SLAC National Accelerator Laboratory, 2575 Sand Hill Road, Menlo Park, California 94025, USA}

\author{J. A. Sobota}
\affiliation{Stanford Institute for Materials and Energy Sciences, SLAC National Accelerator Laboratory, 2575 Sand Hill Road, Menlo Park, California 94025, USA}

\author{Z.-X. Shen}
\affiliation{Stanford Institute for Materials and Energy Sciences, SLAC National Accelerator Laboratory, 2575 Sand Hill Road, Menlo Park, California 94025, USA}
\affiliation{Geballe Laboratory for Advanced Materials, Department of Physics and Applied Physics, Stanford University, Stanford, California 94305, USA}

\date{\today}

\begin{abstract}
We report time- and angle-resolved photoemission spectroscopy measurements on the Sb(111) surface. We observe band- and momentum-dependent binding-energy oscillations in the bulk and surface bands driven by $A_{1g}$ and $E_{g}$ coherent phonons. While the bulk band shows simultaneous $A_{1g}$ and $E_{g}$ oscillations, the surface bands show either $A_{1g}$ or $E_{g}$ oscillations. The observed behavior is reproduced by frozen-phonon calculations based on density-functional theory. This evidences the connection between electron-phonon coupling and coherent binding energy dynamics.   
\end{abstract}

\maketitle

In recent years, there has been growing interest in using non-equilibrium techniques to probe equilibrium material properties. Coherent phonons, which are non-equilibrium atomic motions driven by an ultrafast light pulse, are particularly useful for this purpose, since the oscillatory displacements of the atoms are associated with simultaneous oscillations in the electronic binding energies. As a result, the lattice and electronic dynamics associated with coherent phonons provide direct information on the equilibrium property of electron-phonon coupling \cite{Khan_1984_deformation,Giovannini_2020_direct}.

Time- and angle-resolved photoemission spectroscopy (trARPES) is one of the most powerful methods to study coherent phonons as it can directly monitor the temporal evolution of electronic band structure. Specifically, it can resolve $\Delta\varepsilon_n(k)$, the electronic energy shift as a function of band index $n$ and electron momentum $k$, separately for each phonon mode. This is proportional to the deformation potential $D_n(k) = \Delta \varepsilon_n(k) / \Delta r$, where $\Delta r$ is the corresponding lattice distortion, which represents the strength of electron-phonon coupling with $n$-, $k$-, and mode-specificity \cite{Khan_1984_deformation}. This technique has been applied to deduce the behavior of electron-phonon coupling in materials with surface states \cite{Papalazarou:2012aa, Faure:2013aa,Sobota:2014aa,Golias:2016aa}, strong electron correlations \cite{Rettig_2015_ultrafast, Gerber:2017aa,Yang_2019_mode}, coexisting phases \cite{Suzuki_2021_detecting}, and complex multi-band electronic structures \cite{Hein_2020_mode}. Integration with ultrafast structural probes to measure $\Delta r$ enables theory-free quantification of the deformation potential \cite{Rettig_2015_ultrafast,Gerber:2017aa}.

As trARPES investigations of coherent phonons advance towards increasingly complex material systems, it is critical to verify that the non-equilibrium probe is faithful to the equilibrium quantity of interest, especially since deviations from expected behavior are taken as evidence of non-trivial physics \cite{Gerber:2017aa}. trARPES experiments on semimetals and topological insulators have shown that frozen-phonon density functional theory (DFT) calculations provide an adequate description of the band- \cite{Faure:2013aa} and $k$-dependence \cite{Papalazarou:2012aa,Golias:2016aa} of binding-energy dynamics attributed to fully-symmetric $A_{1g}$ coherent phonons. It is desirable to extend this analysis to modes of different symmetries, preferably in a system which exhibits a band- and $k$- dependent response, to establish a comprehensive benchmark across the parameter space relevant to electron-phonon coupling in complex materials. 

\begin{figure*}
\begin{center}
\includegraphics[width=17.5 cm]{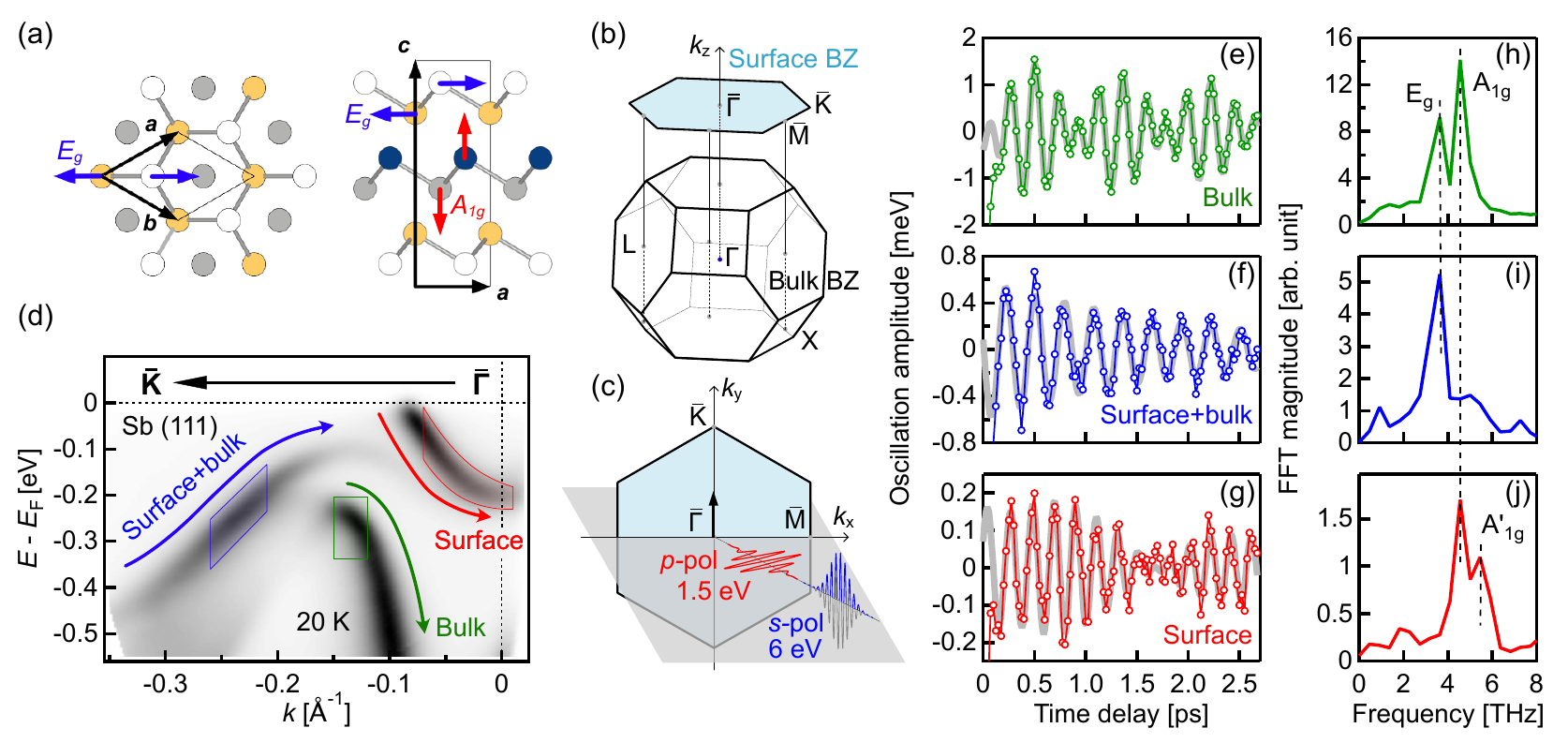}
\caption{(a) Top and side views of the crystal structure of Sb. Blue and red arrows represent the atom displacement for the $A_{1g}$ and $E_{g}$ phonons. (b) Bulk and surface Brillouin zone of Sb. (c) Experimental geometry. (d) Equilibrium experimental ARPES spectrum along the $\overline{\Gamma}$ - $\overline{\rm K}$ direction. Coherent phonon induced binding-energy oscillations and the Fourier power spectra for the bulk band ((e) and (h)), the surface+bulk band ((f) and (i)), and the surface band ((g) and (j)), marked by green, blue, and red arrows in panel (d), respectively. The gray curves in panels (e), (f), and (g) are fitted curves.}
\label{Fig1}
\end{center}
\end{figure*}

Sb is an ideal material for such a study. Sb is a topological semimetal \cite{Hsieh:2009aa,Seo:2010aa,Zhang:2012aa} with bulk and surface bands well-described by DFT \cite{Bian_2011_passage} and accessible by photoemission with laser sources \cite{Xie:2014aa}. 
Sb has a rhombohedral A7 crystal structure (Fig.~\ref{Fig1}(a)), which is a cubic lattice distorted along the (111) direction (or the $c$-axis direction in a hexagonal representation). The distortion happens due to a Peierls instability along the (111) direction, and Sb atoms form honeycomb-like bilayers.
This structure hosts a total of two optical phonon modes ($A_{1g}$ and $E_{g}$), both of which are susceptible to coherent excitation \cite{Cheng:1990aa,Ishioka:2008aa}, and the mechanism of which has been studied intensively \cite{Zeiger:1992aa,Garrett:1996aa,Stevens:2002aa,Shinohara:2012aa,Campi:2012aa}.

This letter reports trARPES measurements on the Sb(111) surface. We observe coherent phonon-induced binding-energy oscillations depending on momentum, band index, and phonon mode, highlighting the interplay of lattice and electronic degrees of freedom. We show that frozen-phonon DFT calculations can qualitatively reproduce the observed behavior, thereby reaffirming that the dynamics of electronic states modulated by coherent phonons are well described by the equilibrium concept of electron-phonon coupling.

Our trARPES setup is based on a Ti:Sapphire regenerative amplifier outputting 1.5~eV, 35~fs pulses at a repetition rate of 312~kHz \cite{Gauthier:2020aa}. 
The photon energy was quadrupled to 6.0~eV for the probe pulse by two stages of second harmonic generation.
The beam profiles for the pump and probe pulses were $68 \times 85$ ${\rm \mu m^2}$ and $38 \times 41$ ${\rm \mu m^2}$ in full width at half maximum, respectively. The fluence of the incident 1.5~eV pump was 0.17~mJ/cm$^2$. 
Photoelectrons were collected by a hemispherical analyzer and spectra were recorded as a function of  pump-probe delay.
The overall time resolution was deduced to be 85~fs from cross correlations of pump and probe pulses. 
The measurement temperature was 20 K.
The light incidence plane was along the mirror plane of the sample, and the pump and probe light polarizations were $p$ and $s$, respectively, as shown in Fig.~\ref{Fig1}(c). Photoelectrons are collected along the $\overline{\Gamma} - \overline{\rm K}$ direction of the surface Brillouin zone as shown by a black arrow in Fig.~\ref{Fig1}(c). To detect weak coherent phonon oscillations, our accumulated data required correction of systematic drifts along the energy, momentum, and time axes as described in the supplementary materials \cite{SOM}.

First-principle calculations were performed on a 9 Sb bi-layer slab (18 Sb layers) with 30~\AA\  vacuum layer using the full-potential augmented-plane-wave method as implemented in the WIEN2k code \cite{Blaha:2001aa}. Note that Sb bilayers become topological with 8 or more bilayers according to a previous DFT calculation \cite{Zhang:2012aa}. The experimental lattice structure was used for the calculation. For the exchange-correlation potential, the generalized gradient approximation (GGA) of Perdew-Burke-Erzerhof parametrization \cite{Perdew:1996aa} was employed with the spin-orbit interaction taken into account. The Brillouin-zone integration was performed on a 20 $\times$ 20 $\times$ 1 $k$-point mesh. We displaced Sb atoms by  $\pm 0.02$, $\pm 0.05$, and $\pm 0.1$\% of the $c$-axis lattice constant (11.22~\AA)\  along the trigonal axis for the $A_{1g}$ phonon and by $\pm 0.01$, $\pm 0.02$, and $\pm 0.05$\% perpendicular to the trigonal axis for the $E_{g}$ phonon. 
These displacement values result in binding energy shifts that are resolvable while maintaining a linear relationship between the energy shift and the displacement  \cite{SOM}. The displacement directions for the $A_{1g}$ and the $E_{g}$ phonon are depicted by red and blue arrows in Fig.~\ref{Fig1}(a), respectively. 
The band structures were calculated for each displacement, and the obtained binding-energy shift ($\Delta \varepsilon_n(k)$) as a function of atom displacement ($\Delta r$) was fitted by a linear function at each momentum to obtain the proportionality constant $\Delta \varepsilon$/$\Delta r$, which corresponds to the deformational potential. In this way, we were able to minimize and characterize errors from the DFT calculations \cite{SOM}.

Figure \ref{Fig1}(d) shows the equilibrium ARPES spectrum taken along the $\overline{\Gamma} - \overline{\rm K}$ direction. 
The spectrum is consistent with previous studies \cite{Sugawara:2006aa, Xie:2014aa} and has three sharp energy bands marked by arrows in Fig.~\ref{Fig1}(d). The band marked by a green arrow is a bulk band, while the band marked by a red arrow is a surface band. The band marked by a blue arrow has surface character near $\overline{\Gamma}$ but has increasing bulk character as $k$ increases (see supplementary materials for the orbital character of each band \cite{SOM}). We thus refer to these three bands as the bulk band (green arrow), the surface band (red arrow), and the surface+bulk band (blue arrow), hereafter.

In order to examine the temporal evolution of the energy bands, we track the binding energy of each band by fitting a Gaussian function to the energy distribution curve (EDC) at each $k$-point and at each delay time. Fig.~\ref{Fig1}(e)-(g) show how the three bands oscillate in binding energy as a function of delay time after the pump pulse. Here, fifth order polynomial backgrounds are subtracted to extract the oscillatory components. For this figure, the oscillatory curves are averaged from $k = - 0.15$ to $-0.12$~\AA$^{-1}$ for the bulk band, $k = - 0.26$ to $-0.21$~\AA$^{-1}$ for the surface+bulk band, and $k = - 0.07$ to $0.01$~\AA$^{-1}$ for the surface band. These integration regions are indicated by boxes in Fig. \ref{Fig1}(d).
The bulk band shows the strongest average oscillation with an amplitude $>1$~meV. The surface+bulk band shows weaker oscillation than the bulk, and the surface band shows the weakest oscillation with an amplitude $<0.2$~meV. The weaker responses of the surface-related bands indicate that the electron-phonon coupling is weaker for the surface bands, as also suggested in Ref.~\cite{Xie:2014aa}.

Figures \ref{Fig1}(h), \ref{Fig1}(i), and \ref{Fig1}(j) show the magnitude of the Fourier transforms of the curves shown in Figs. \ref{Fig1}(e), \ref{Fig1}(f) and \ref{Fig1}(g).
The Fourier transform of the bulk-band oscillation has two peaks around 3.6~THz and 4.5~THz, which correspond to the $E_{g}$ and $A_{1g}$ phonon modes \cite{Wang:2006aa}, respectively. The multi-frequency oscillation can also be seen as a beating pattern in Fig.~\ref{Fig1}(e). The surface+bulk band does not show $A_{1g}$ oscillations but shows $E_{g}$ oscillation only. On the contrary, the surface band does not couple to the $E_{g}$ phonon but couples to the $A_{1g}$ phonon.

The surface band has an additional higher-frequency mode around 5.5 THz, which has not been observed experimentally thus far to our knowledge but was predicted theoretically as a stiffening of the surface bilayer with respect to the bulk \cite{Campi:2012aa}. We refer to this higher-frequency mode as the $A_{1g}'$ mode.
Our results corroborate association of the $A_{1g}'$ mode with the surface because it is only present in the surface band and was absent in  previous bulk-sensitive Raman spectroscopy \cite{Wang:2006aa} and time-resolved reflectivity (TRR) measurements \cite{Ishioka:2008aa}.
A previous trARPES study \cite{Sobota:2014aa} reported that Bi$_{2}$Se$_{3}$ also shows a mode associated with the surface state, the frequency of which is lower than that of the bulk $A_{1g}$ mode. The opposite sign of the effect in these two materials suggests a difference in the nature of their interlayer atomic forces.

\begin{table}
\caption{Fitting parameters for Eq. \ref{fitting}.}
\label{FitTable}
\centering
\begin{tabular}{lccccc}
\hline\hline
 & \multicolumn{2}{c}{Bulk} & \multicolumn{1}{c}{Surface+bulk} & \multicolumn{2}{c}{Surface}\\
  & $A_{1g}$ & $E_{g}$ & $E_{g}$ & $A_{1g}$ & $A_{1g}'$ \\
  \hline
$f$ [THz]~~~   & 4.66(1) & 3.49(1)& 3.50(1) & 4.66(2) & 5.25(3)\\

$A$  [meV]\   & 0.89(4) & 0.69(4) & 0.60(5) & 0.23(3) & 0.04(1)\\

$\phi$ [$\pi$]\   & -0.68(2) &  0.46(2) & 0.45(3) & -0.41(4) & 0.56(8)\\

$1/\tau$  [/ps]\   & 0.13(3)  &  0.23(4)& 0.48(7) & 0.8(2) & -0.2(2)\\

\hline\hline

\end{tabular}
\end{table}

To be more quantitative, we perform a curve fit using two cosine functions with exponential decay, as shown below.
\begin{equation}
\begin{split}
 \Delta E = &A_{1}\cos (2 \pi f_{1} t + \phi_{1}) \exp (-t/\tau_{1}) \\
  + &A_{2}\cos (2 \pi f_{2} t + \phi_{2}) \exp (-t/\tau_{2}).
\end{split}
\label{fitting}
\end{equation}

Here, $\Delta E$ denotes the shift of the binding energy, $f_{1,2}$ and $\phi_{1,2}$ denote the frequency and the phase of the oscillation, $\tau_{1,2}$ represents the decay time. The fits are represented by gray curves in Figs. \ref{Fig1}(e), \ref{Fig1}(f), and \ref{Fig1}(g), and they reproduce the data well. The deduced fitting parameters are summarized in Table \ref{FitTable}. 

\begin{figure}[t!]
\begin{center}
\includegraphics[width=8.2 cm]{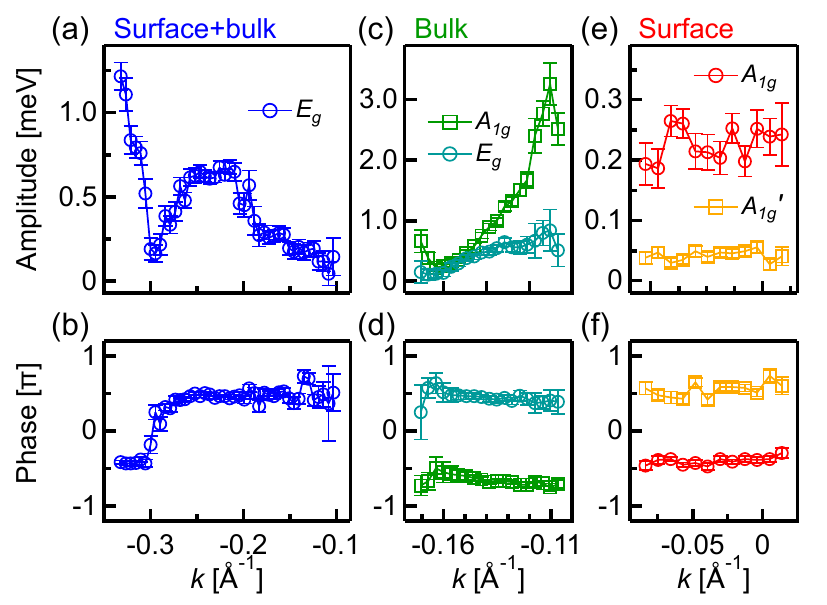}
\caption{
Momentum dependence of the binding-energy oscillation amplitudes (a) and phases (b) for the surface+bulk band, the bulk band (c) and (d), and the surface band (e) and (f).
}
\label{Fig2}
\end{center}
\end{figure}

The fitted frequencies of the $A_{1g}$ and $E_{g}$ phonon modes are $4.66\pm0.01$ and $3.49\pm0.01$~THz, consistent with the frequencies of 4.65 and 3.47~THz observed in TRR measurements \cite{Ishioka:2008aa}. The decay rates of $A_{1g}$ and $E_{g}$ phonons in the bulk band are $0.13 \pm 0.03$ and $0.23 \pm 0.04$ ps$^{-1}$, also comparable to the decay rates of 0.092 and 0.31 ps$^{-1}$ observed in the TRR measurements. Although the bulk band behaves consistently with the TRR measurement, the surface band and the surface+bulk band show faster decay, possibly suggesting increased dampening near the surface.

Figure \ref{Fig2} shows the momentum dependence of the band oscillation amplitudes and phases. Here, Eq. \ref{fitting} was fitted to the EDC peak-position oscillation at each momentum with the decay rates and the frequencies fixed to the ones shown in Table \ref{FitTable} to minimize the number of free parameters. 
The surface+bulk band shows peculiar behavior: the phase rotates by $\pi$ at $k = -0.3$~\AA$^{-1}$. This behavior is reminiscent of anti-phase oscillations reported in Bi$_{2}$Te$_{3}$ \cite{Golias:2016aa} and BaFe$_{2}$As$_{2}$ \cite{Okazaki:2018aa}. The present finding differs in that the pivoting occurs at a seemingly arbitrary $k$-point, and is not associated with high-symmetry directions in the Brillouin zone.

In contrast, the bulk band and the surface band exhibit nearly constant phases. The bulk-band oscillations increase in amplitude approaching the $\overline{\Gamma}$ point (Fig. \ref{Fig2}(c)), while the surface band oscillations show little momentum dependence in the measured range (Figs. \ref{Fig2}(e) and \ref{Fig2}(f)).

\begin{figure}[t!]
\begin{center}
\includegraphics[width=8.2 cm]{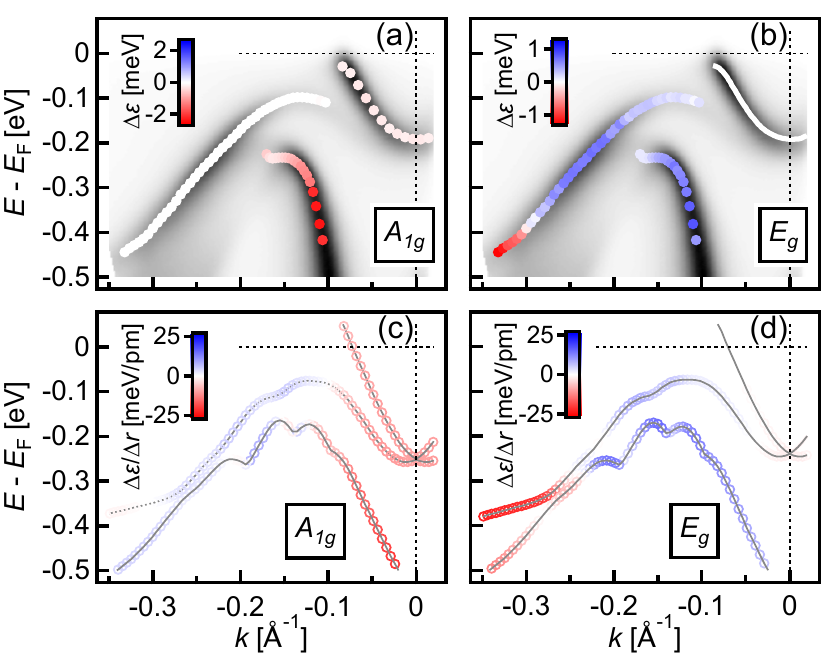}
\caption{$A_{1g}$ and $E_{g}$ phonon oscillation amplitudes on each band at each momentum obtained in the experiments (a) and (b), and the DFT-calculated deformation potentials (c) and (d). The oscillation amplitudes are represented by color.}
\label{DFT}
\end{center}
\end{figure}

Figures \ref{DFT}(a) and \ref{DFT}(b) visualize the momentum- and band-dependent oscillation amplitude for the $A_{1g}$ and $E_{g}$ phonon modes. Filled circles are plotted at the EDC peak positions, with their colors representing the signed oscillation amplitudes determined by multiplication with a phase factor, namely $\Delta \varepsilon(k) = A(k) \times \sin(\phi(k))$. It can be clearly seen that the surface+bulk band reverses oscillation phase at $k \sim -0.3$~\AA$^{-1}$. Because the $E_{g}$ mode was not detected for the surface band, we use white solid markers for its peak position in Fig.~\ref{DFT}(b).

To understand the momentum dependence of oscillation amplitudes and phases, we performed frozen-phonon DFT calculations and compare them with the experimental data. Figures \ref{DFT}(c) and \ref{DFT}(d) show the calculated deformation potential $\Delta \varepsilon$/$\Delta r$, plotted similarly to the Figs. \ref{DFT}(a) and \ref{DFT}(b). 
DFT calculates a non-zero coupling for all modes and bands, and thus the absence of certain modes in experiment does not signify that these mode couplings are symmetry-forbidden \cite{SOM}.

For the $A_{1g}$ mode, the surface+bulk band shows weak but finite response. In contrast to the experimental results, the surface and bulk bands show similar magnitudes of response; we speculate that the weak surface response in experiment may be attributed to a smaller surface bilayer distortion as compared to the deeper layers because the surface bilayer is stiffer \cite{Campi:2012aa}. This speculation should be tested in future studies by directly measuring the atomic motion using time-resolved diffraction techniques \cite{Greif_2016,Gerber:2017aa,Waldecker_2017}, although it may be challenging to separately detect surface atom motion.

On the other hand, for the $E_{g}$ mode, the calculation well reproduced the experimental observations: the surface+bulk band indeed shows a phase reversal around $k = -0.25$ \AA$^{-1}$, and the surface band does not respond to the $E_{g}$ displacement.
The improved agreement for the $E_{g}$ mode compared to the $A_{1g}$ mode may be attributed to the fact that the $E_{g}$ distortion is less sensitive to the surface termination since its displacement direction is perpendicular to the surface. Despite these minor discrepancies, we find that the experimental data is qualitatively well described by the frozen phonon DFT calculations. It is worth noting that the $E_{g}$ phase reversal occurs where the two bands approach each other, and therefore the reversal may be associated with their hybridization in this region. 

In summary, the present study has revealed band, momentum, and phonon-mode-dependent electron-phonon coupling in Sb(111), which have been well reproduced by density-functional-theory calculations. It has been demonstrated that coherent phonons do not only rigidly shift bands in energy, but also exhibit a dependence on bulk/surface character as well as interband hybridizations. The fact that these behaviors are captured in frozen-phonon DFT calculations provides strong evidence that coherent phonon responses are rooted in the equilibrium concept of electron-phonon coupling. These results further justify the use of trARPES to investigate strongly-correlated materials, in which the electron-phonon interactions are intrinsically intertwined with the effect of strong electron interactions.

\newpage
\section{Supplementary Information}
\subsection{Data correction to account for drifts in the measurement}

In this section, the details of the data analysis are explained. Fig.~\ref{drift}(a) shows the angle-resolved photoemission spectroscopy (ARPES) spectra before the angle to momentum conversion. In the present study, in order to detect weak sub-meV binding energy oscillations, we accumulated the data for nearly two days. The accumulated dataset consists of $\sim$300 iterations of delay scans. During this substantial integration time, the ARPES spectra exhibited the following drifts:

\begin{figure*}[b]
\begin{center}
\includegraphics[width=16.0 cm]{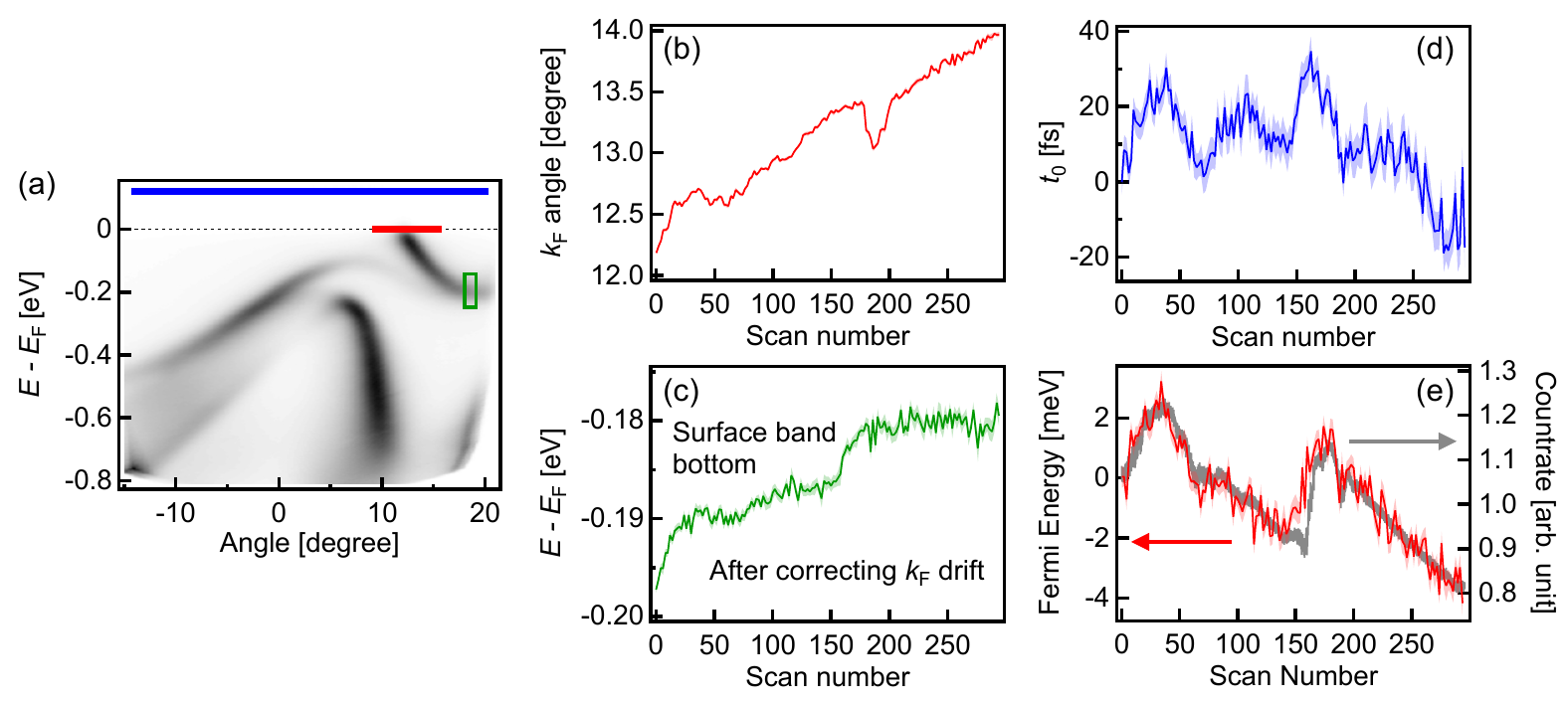}
\caption{(a) Angle-resolved photoemission spectra. Scan number dependence of (b) $k_{\rm F}$ angle, (c) energy position of the surface band bottom indicated by the green box in panel (a), (d) time zero ($t_{0}$) estimated from the intensities above the Fermi level indicated by the blue line in panel (a), and (e) Fermi energy and photoelectron countrate. Note that each ``scan'' consists of a complete delay scan, so that $t_0$ can be determined independently for each iteration.}
\label{drift}
\end{center}
\end{figure*}

\begin{itemize}

    \item  $k_{\rm F}$ drift (Fig. \ref{drift}(b)): the angle corresponding to the Fermi wave number $k_{\rm F}$, deduced from the region indicated by the red line in Fig. \ref{drift}(a), increased by $\sim 2$ degrees. This is likely related to long-term mechanical instabilities of the sample manipulator and/or laser beam pointing drifts.

\item $E_{\rm B}$ drift (Fig. \ref{drift}(c)): the binding energy of the surface band bottom, indicated by the green box in Fig. \ref{drift}(a), increased by $\sim 20$~meV.
This is probably caused by a cumulative gas absorption on the Sb surface, which can shift the chemical potential.

\item $t_{0}$ drift (Fig. \ref{drift}(d)): time zero ($t_{0}$), at which both pump and probe laser pulses are incident on the sample simultaneously, varied within $\sim 50$ fs. $t_{0}$ was deduced from the step-function-like rise in the intensities above the Fermi level indicated by the blue line in Fig. \ref{drift}(a). This $t_{0}$ drift is attributed to optical path length changes due to temperature variations of the laboratory.

\item $E_{\rm F}$ drift (Fig. \ref{drift}(e)): the Fermi energy drift of $\sim 6$ meV was rather subtle compared to the $E_{\rm B}$ drift. This weak $E_{\rm F}$ drift likely resulted from a change of space-charging due to laser intensity variations, as the $E_{\rm F}$ shift tracks the photoelectron countrate.

\end{itemize}

\begin{figure*}
\begin{center}
\includegraphics[width=16.3cm]{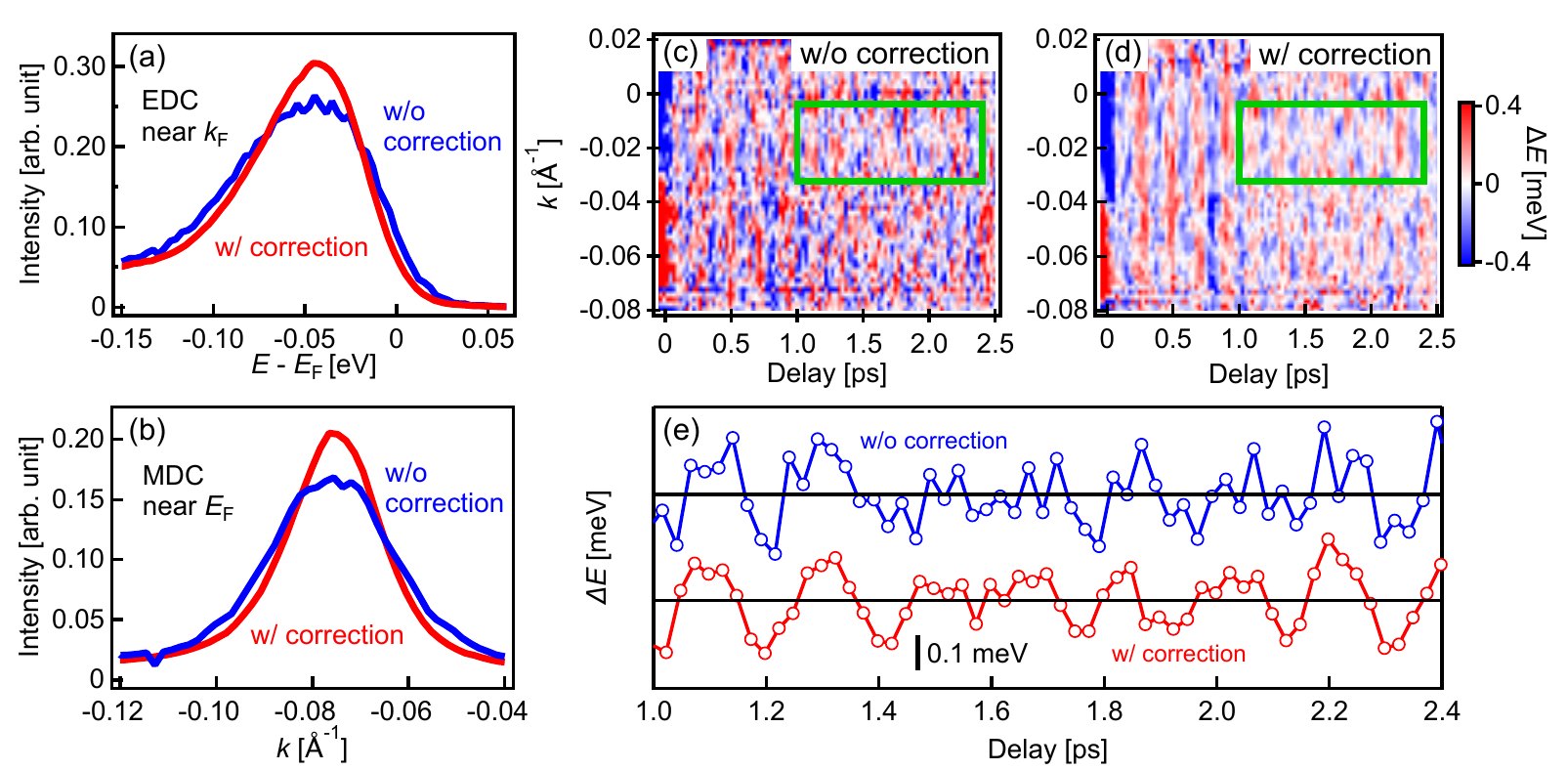}
\caption{(a) Energy distribution curve (EDC) at $k_{\rm F}$ and (b) Momentum distribution curve (MDC) at $E_{\rm F}$ with and without the drift correction. (c), (d) Momentum-dependent binding energy oscillation due to the pump excitation with or without the drift correction. (e) Averaged binding energy oscillation in the green boxes indicated in panels (c) and (d).}
\label{correction}
\end{center}
\end{figure*}

If the trARPES spectra are simply added to each other, the above-mentioned drifts effectively broaden the distributions along the momentum, energy and time axes. This degradation of data quality is particularly detrimental for coherent phonon studies in which all three variables are fundamentally coupled. To overcome this issue, we individually shift each scan in angle, energy, and delay, in that order, to compensate for the $k_{\rm F}$, $E_{\rm B}$, and $t_{0}$ drifts. Figs.~\ref{correction}(a) and \ref{correction}(b) show the energy distribution curve (EDC) at $k_{\rm F}$ and the momentum distribution curve (MDC) at $E_{\rm F}$ with or without the drift corrections.
With the drift corrections, both the MDC and EDC becomes about 35\% sharper in the full width half maximum. 

Figs.~\ref{correction}(c) and \ref{correction}(d) show the pump-induced binding energy oscillation for the surface band with and without drift corrections. Here, the vertical axis represents the wave vector along the $\Gamma$-K direction, and the horizontal axis is the pump-probe delay. Blue and red colors represent the binding energy shift, and the stripe pattern indicates that the surface band moves up and down in energy periodically. Oscillations are more discernible in the data after correction. The oscillations in the green boxes in Figs. \ref{correction}(c) and \ref{correction}(d) are averaged and shown in Fig. \ref{correction}(e). The drift correction clarified the binding energy oscillations considerably, even in the range of 1.4 to 2.4 ps, where the oscillatory signal was weak without the drift correction.

\subsection{Atomic and orbital characters of the energy bands }
\begin{figure*}
\begin{center}
\includegraphics[width=16.3cm]{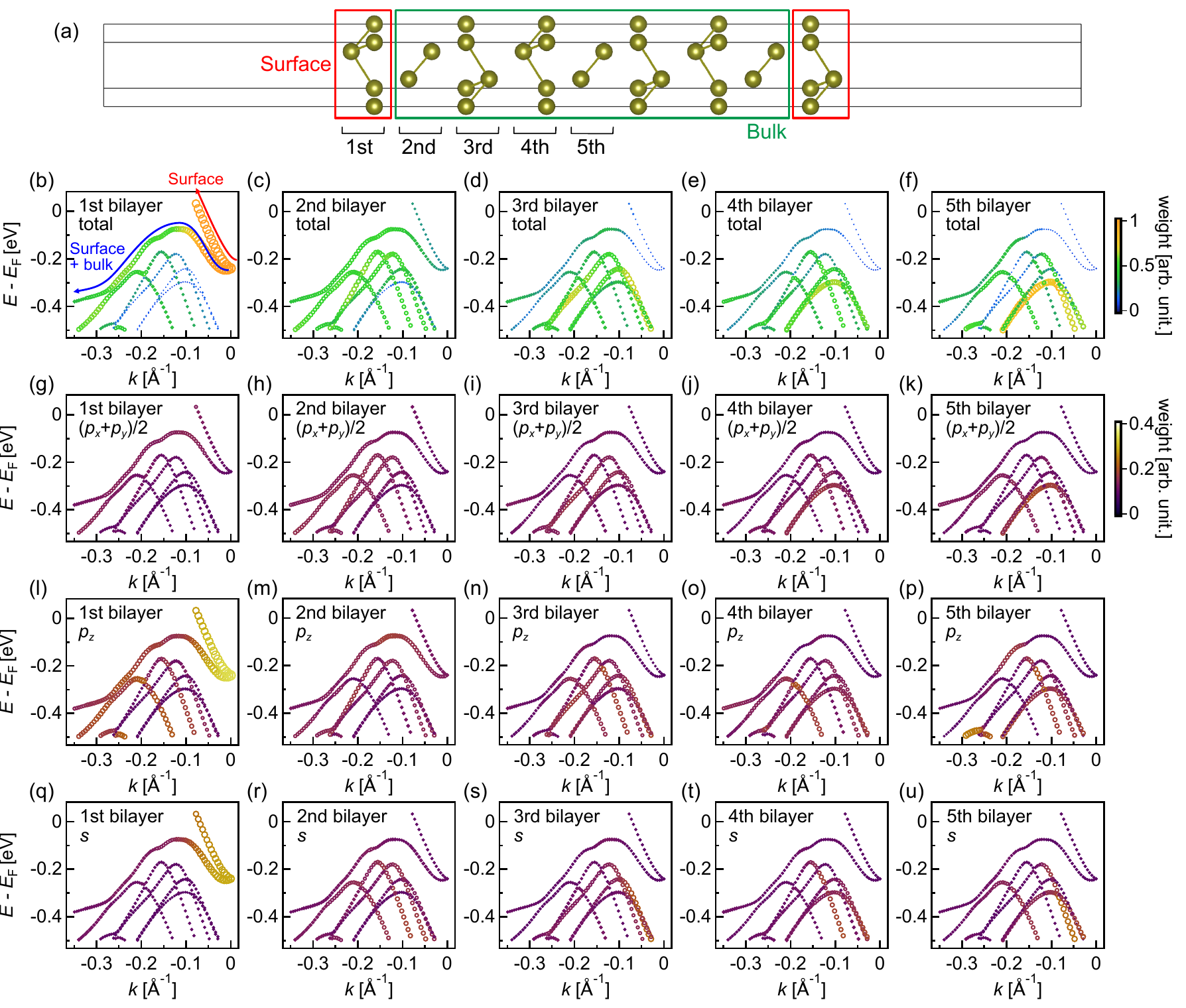}
\caption{(a) 9 Sb bilayers slab used for the calculations. (b)-(f) Sb bilayer weights for each band. (g)-(u) Orbital characters of each band. }
\label{character}
\end{center}
\end{figure*}

In this section, the atomic and orbital characters of the energy bands along the $\Gamma$-K direction near the Fermi level are shown. 
Fig.~\ref{character}(a) shows the slab consisting of 9 Sb bilayers, which is used for the density functional theory (DFT) calculations.
Figs.~\ref{character}(b)-(f) show the weight of each Sb bilayer. 
The band that crosses the Fermi level predominantly originates from the surface Sb bilayer. Therefore, we refer to this band as the surface band in the main text.
The band marked by the blue arrow in Fig. \ref{character}(b) is mainly localized in the 1st and 2nd Sb bilayers, and the weight in the 2nd Sb bilayer becomes relatively strong for $k > 0.1$ \AA$^{-1}$. The 3rd and 5th Sb bilayers also have considerable weight in the regions of $0.1 < |k| < 0.15$ \AA$^{-1}$ and $|k| > 0.15$ \AA$^{-1}$, respectively, as shown in Fig. \ref{character}(d) and \ref{character}(f). Thus, we refer to this band as the surface+bulk band in the main text.

Figs.~\ref{character}(g)-(u)
further decompose the weights into $p_{xy}$, $p_{z}$, and $s$ orbitals. 
The surface band predominantly consists of $p_{z}$ and $s$ orbitals of the first Sb bilayer. The remaining bands contain all orbital characters to some extent, indicating that the Sb $s$ and $p$ orbitals hybridize with each other in the bulk.

\subsection{Extraction of deformation potential}

\begin{figure*}
\begin{center}
\includegraphics[width=16.5cm]{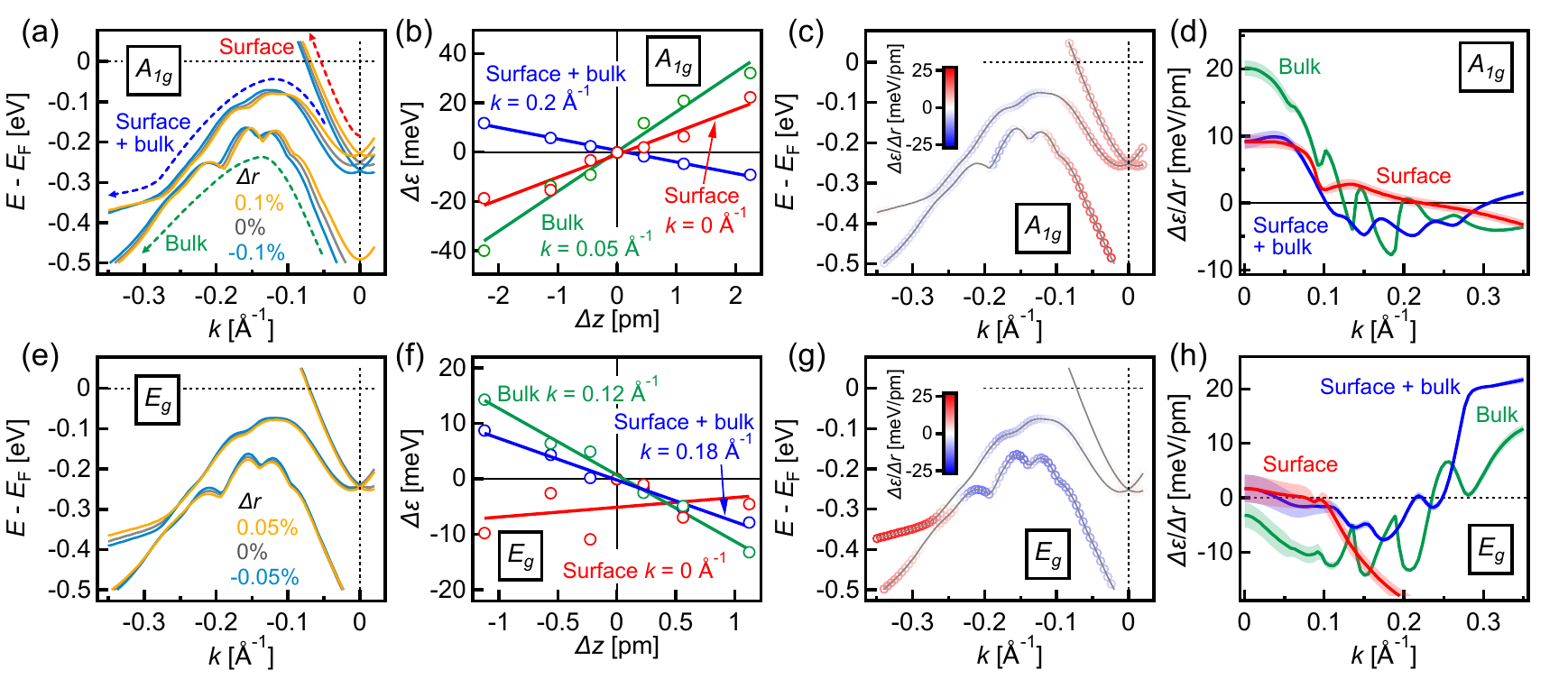}
\caption{Sb band structure and its changes due to $A_{1g}$ (a) and $E_{g}$ (e) atomic displacements. (b) and (f) show band shifts at selected $k$-points for each band as function of displacement. The solid lines are fits by a linear function. (c) and (g) visualize the deformation potentials deduced by the linear fitting. (d) and (h) The $k$-dependent deformation potentials for each band with shaded area representing the fitting errors.}
\label{DFT_S4}
\end{center}
\end{figure*}

In this section, we explain how we extracted the deformation potential ($D = \Delta\varepsilon/\Delta r$), i.e. the energy shift ($\Delta\varepsilon$) caused by atom displacement ($\Delta r$), from the density functional theory (DFT) calculations. Fig.~\ref{DFT_S4}(a) shows the calculated band structures with and without the $A_{1g}$ atomic displacements $\Delta r$ of $\pm 0.1$\% of the c-axis lattice constant ($\pm 1.12$ pm). 
Here, the $A_{1g}$ displacement of $+\Delta r$ means that the interlayer distance within a Sb bilayer is elongated by $+\Delta r$. Binding energies change systematically and monotonically with atomic displacements. The binding energy shifts of each band as a function of $A_{1g}$ displacement are plotted in Fig. \ref{DFT_S4}(b). The  shifts were fitted by a linear function.
The slope represents the deformation potential and is indicated in Fig. \ref{DFT_S4}(c) by color. This figure is the same as the one shown in the main text. 
Fig.~\ref{DFT_S4}(d) also shows the band- and momentum-dependent deformation potentials with shaded area representing error bars from the linear fits. The fitting errors of the deformation potentials are less than 1.5 meV/pm.

The same calculations and analyses are performed for the $E_{g}$ displacements, and the results are shown in Figs. \ref{DFT_S4}(e)-(h). Compared to the $A_{1g}$ case, the band shifts in Fig.~\ref{DFT_S4}(f) seem scattered, especially near the $\Gamma$ point, which is reflected in Fig.~\ref{DFT_S4}(h) by the increased fitting errors up to 2.5 meV/pm. 
This inaccuracy likely results from the fact that the $E_g$ displacement lowers the crystallographic symmetry which makes the DFT calculations more difficult to converge. 
This method allows us to estimate uncertainties due to the inaccuracy of DFT calculations.

\subsection{Electron-phonon couplings allowed and forbidden by symmetry}

Group theory is useful to establish which electron-phonon couplings are allowed or forbidden based on symmetry. The electron-phonon coupling is characterized by the deformation potential $D_n(k)$, which corresponds to the matrix element $\left< \Psi_n(k) \right| H_{el-ph} \left|\Psi_n(k)\right>$. Here, $\left|\Psi_n(k)\right>$ is the electronic state $n$ at momentum $k$ and $H_{el-ph} $ is the electron-phonon coupling Hamiltonian, which is described by the same irreducible representation as the phonon considered. Group theory is used to evaluate if this matrix element is exactly zero (forbidden coupling) or can be finite (allowed coupling) for defined phonon symmetries and electronic state symmetries. This is done by determining if $H_{el-ph} \left|\Psi_n(k)\right>$ is orthogonal or not to $\left< \Psi_n(k) \right|$. For example, if we consider the coupling to an $A_{1g}$ phonon, $H_{el-ph}$ is described by the trivial $A_{1g}$ irreducible representation. The term $H_{el-ph} \left|\Psi_n(k)\right>$ is therefore characterized by ${A_{1g}} \otimes \Gamma_{\Psi_n(k)}=\Gamma_{\Psi_n(k)}$, where $\Gamma_{\Psi_n(k)}$ is the electronic state irreducible representation. As $H_{el-ph} \left|\Psi_n(k)\right>$ and $\left< \Psi_n(k) \right|$ have the same symmetry, they are not orthogonal. Consequently, electron-phonon coupling of $A_{1g}$ phonons is allowed with electronic states of any symmetry. This conclusion for $A_{1g}$ phonons at $Q=0$ is not limited to Sb and applies generally.

While the case for $A_{1g}$ phonons is generic, the electronic state symmetry must be considered explicitly to evaluate the coupling to the $E_g$ phonon in Sb. Neglecting spin-orbit coupling, the Sb electronic states at $k=0$ are described by the $D_{3d}$ point group. In that case, it can be shown that $E_g$ phonons are allowed to couple to $p_x$ and $p_y$ orbitals but coupling to $s$ and $p_z$ orbitals is forbidden. However, coupling to all orbitals is allowed if spin-orbit coupling is present, which is the case in Sb. Considering electronic states with finite momentum ($k\neq0$), we also find that coupling to all orbitals is allowed. Consequently, electron-phonon coupling of $A_{1g}$ and $E_g$ phonons with any electronic states is allowed in Sb. This is in agreement with the DFT that indicates non-zero coupling for all modes and bands.

\subsection{ACKNOWLEDGMENTS}
This work was supported by the Department of Energy, Office of Basic Energy Sciences, Division of Materials Science and Engineering. S.S. acknowledges financial support from the JSPS Research Fellowship for Research Abroad.

\bibliography{Bibtex_all.bib}

\end{document}